\documentclass[11pt]{article}

\usepackage{amsfonts}
\usepackage{amsmath}
\usepackage{amsthm}
\usepackage[british]{babel}
\usepackage[a4paper]{geometry}
\usepackage{graphicx}
\usepackage{subcaption}
\usepackage{hyperref}

\newtheorem{theorem}{Theorem}[section]

\newtheorem{proposition}[theorem]{Proposition}

\theoremstyle{definition}

\theoremstyle{remark}

\newtheorem{remarkx}[theorem]{Remark}
\newenvironment{remark}
{\pushQED{\qed}\remarkx}
{\popQED\endremarkx}
\newcommand{\E}{\mathbb E}
\newcommand{\N}{\mathbb N}
\renewcommand{\P}{\mathbb P}
\newcommand{\R}{\mathbb R}
\newcommand{\Z}{\mathbb Z}
\newcommand{\sgn}{\operatorname{sgn}}
\title{Scaling Results for Piecewise Deterministic Monte Carlo : A Survey}

\author{Joris Bierkens \thanks{Delft Institute of Applied Mathematics, \href{mailto:joris.bierkens@tudelft.nl}{joris.bierkens@tudelft.nl}}}

\begin{document}
\maketitle

\section{Introduction}
In the past decade the notion of Piecewise Deterministic Monte Carlo (PDMC) was uncovered by the Bayesian computation community as a promising alternative of classical discrete time MCMC algorithms \cite{BouchardCoteVollmerDoucet2017,BierkensRoberts2017}. The underlying methodology had by then already been applied successfully for computations in statistical physics (albeit using a different terminology) \cite{MichelKapferKrauth2014,PetersDeWith2012} and had also been explored to some extent by the applied probability community \cite{micloEtudeSpectraleMinutieuse2013,monmarcheHypocoerciveRelaxationEquilibrium2014,Monmarche2016}.

PDMC methods are built upon continuous time stochastic processes known as Piecewise Deterministic Markov Processes (PDMPs) \cite{Davis1984, Davis1993}. As the name indicates, these processes move along deterministic trajectories until at random time the process changes position or direction, in such a way that the resulting process is Markov.
These new sampling methods led to a surge of research interest. Examples of PDMC algorithms are the Bouncy Particle Sampler \cite{BouchardCoteVollmerDoucet2017}, the Zig-Zag Process \cite{BierkensFearnheadRoberts2016}, the Coordinate Sampler \cite{WuRobert2018}, Event Chain Monte Carlo \cite{MichelKapferKrauth2014,shibaDiffusiveScalingLimits2026} and Randomized Hamiltonian Monte Carlo \cite{bou-rabeeRandomizedHamiltonianMonte2017}.

Several papers have sought to develop PDMC methodology dedicated to specific application areas, for example of target distributions with atoms as in Bayesian spike and slab regression \cite{bierkensStickyPDMPSamplers2022}, or discontinuous target distributions \cite{bierkensMethodsApplicationsPDMP2023}. One of the key properties of PDMC methods is principled subsampling, which enables data subsampling \cite{BierkensFearnheadRoberts2016} or other uses of stochastic gradients (e.g. in a Girsanov type change of measure, in the simulation of diffusion processes \cite{bierkensPiecewiseDeterministicMonte2021})  without affecting the correctness of the target distributions.

The efficient simulation of the continuous time trajectory is often non-trivial and several approaches have been developed: for adaptive pre-conditioning \cite{BerBie2022}, and numerical discretization schemes (e.g., following an Euler scheme \cite{BertazziBierkensDobson22}, higher order splitting schemes \cite{JMLR:v26:23-0036} or alternative approaches \cite{corbellaAutomaticZigZagSampling2022,paganiNuZZNumericalZigZag2024}.

Naturally the community was interested in determining convergence rates and results on computational efficiency of PDMC.
In this survey paper we discuss some key results in this direction. After the introduction of two well known PDMC methods (Bouncy Particle Sampler \cite{BouchardCoteVollmerDoucet2017} and the Zig-Zag Process \cite{BierkensRoberts2017}) in Section~\ref{sec:preliminaries}, we consider the ergodic properties of these samplers in Section~\ref{sec:ergodicity}. We then discuss Central Limit Theorems for ergodic averages in Section~\ref{sec:CLT}.

One of the highlights of the theory of classical MCMC is the result on optimal scaling with dimension of RWM in \cite{RobertsGelmanGilks1997}. We describe a similar analysis in the context of PDMPs in Section~\ref{sec:dimension}. Another important scaling question arises when we consider sampling from target probability distributions with narrow ridges; this setting is considered in Section~\ref{sec:anisotropy}.

Finally, PDMC methods are able to employ unbiased stochastic gradients of the log target distribution  without affecting correctness of the stationary distribution \cite{BierkensFearnheadRoberts2016}. A detailed analysis of the scaling of the Zig-Zag Process in the big data regime while using random subsampling has been carried out in \cite{agrawalLargeSampleScaling2024} and a brief overview of these results is presented in Section~\ref{sec:big-data}.

\section{Preliminaries}
\label{sec:preliminaries}

In principle there are many ways to design PDMC methods. In order to keep technicalities at a minimum we will only discuss the setting of the Bouncy Particle Sampler (BPS, \cite{BouchardCoteVollmerDoucet2017}) and the Zig-Zag process (ZZP, \cite{BierkensFearnheadRoberts2016}), for two reasons: (i) these processes are in some sense the most basic in their formulation and widely used in applications, and (ii) our scaling results are obtained specifically for these processes.

\subsection{Common structure for ZZP and BPS}

The common objective of ZZP and BPS is to simulate a target distribution $\pi(dx) = \pi(x) \, d x$ on $\R^d$.
Here the density $\pi(x)$ will be expressed as $\pi(x) = \exp(-U(x)) / Z$, where $Z$ is a (possibly unknown) normalization constant.

 We will augment the state space to $E = \R^d \times \mathcal V$, where $\mathcal V$ is a suitable set of velocities; the particular choice of $\mathcal V$ is different for ZZP and BPS. We use $\mathcal E$ to indicate the Borel $\sigma$-algebra on $E$. Typical elements in $E$ are denoted by $z = (x,v)$ with $x \in \R^d$ and $v \in \mathcal V$.

We will define a continuous time stochastic process $Z_t = (X_t, V_t)_{t \ge 0}$ in $E$ with the following properties:
\begin{itemize}
	\item[(i)] With probability one, at almost all times $t \ge 0$, $(X_t, V_t)$ satisfies the relation $\frac{dX_t}{dt} = V_t$, indicating that $V_t$ is the velocity of $X_t$.
	\item[(ii)] At random times, the process changes velocity at a state-dependent rate $\lambda(X_t,V_t)$  according to a Markov transition kernel $Q$. Here $Q(x, v, dv')$ indicates the law of jumping from current state $(x,v)$ to a new velocity $v'$; the position remains unchanged.
	\item[(iii)] The process $(X_t, V_t)$ is a Markov process and has a stationary distribution $\pi(dx,dv)$ with marginal distribution $\pi(dx) = \int_{v \in \mathcal V} \pi(dx,dv)$ as desired for the purposes of Monte Carlo computation.
\end{itemize}

Throughout we assume that $\lambda: E \rightarrow [0,\infty)$ is continuous and $t \mapsto \lambda(X_t,V_t)$ is absolutely continuous.

A mathematical construction of the process first defines the value of a discrete time skeleton process $(\check X_i, \check V_i, T_i)$, from which the continuous time process $(X_t,V_t)$ can be obtained by interpolation.

Let $\nu$ denote a probability distribution over $E$ representing the distribution of the initial state. The skeleton process is constructed as follows:
\begin{itemize}
	\item Draw $(\check X_0, \check V_0) \sim \nu$. Set $T_0 = 0$.
	\item For $i =1,2, 3, \dots$:
	\begin{itemize}
	\item Draw $\tau_i \ge 0$ independent of $(\check X_j, \check V_j)_{j = 0, \dots,  i-1}$, with distribution 
	\[ \P(\tau_i \ge t \mid \check X_{i-1}, \check V_{i-1}) = \exp \left( -\int_0^t \lambda(\check X_{i-1} + s \check V_{i-1} , \check V_{i-1}) \, d s \right).\] 
	\item Set $\check X_i = \check X_{i-1} + \tau_i \check V_{i-1}$ for the new position.
	\item Draw a new velocity $\check V_i$ according to the distribution $Q(\check X_i, \check V_{i-1}, \cdot)$
	\item Set $T_i = T_{i-1} + \tau_i$.
\end{itemize}
\end{itemize}
The obtained values $(\check X_i, \check V_i)$ can be interpreted as the values of the continuous time process at the times $T_i$ respectively, for $i =0, 1, 2, \dots$.
We can then obtain the value of the continuous time process by setting 
\[ X_t = \check X_i + (t-T_i) \check V_i, \quad V_t = \check V_i, \quad \text{for $t \in [T_i, T_{i+1})$}.\]
This completely specifies the construction of the PDMP $(X_t,V_t)_{t \ge 0}$ for a given combination of initial condition $\nu$, jump intensity $\lambda$ and transition kernel $Q$. 

Throughout this survey our choice of $\lambda$ and $Q$ will guarantee that the process is well-defined for all $t \ge 0$. 

Clearly $\lambda$ and $Q$ should be related in some way to $\pi$, in order to guarantee that $Z_t$ has the correct stationary distribution.
We will now describe the particular choice of $\mathcal V$, $\lambda$ and $Q$ for the ZZP and the BPS.

\subsection{Zig-Zag Process}

The Zig-Zag Process (ZZP, \cite{BierkensFearnheadRoberts2016}) has a discrete set of velocities, $\mathcal V = \{ -1,  1\}^d$, i.e., in every coordinate direction, the process moves either with velocity $+1$ or $-1$.

The velocity switching mechanism is easiest understood as having one jump rate $\lambda_i$ for each coordinate $i = 1, \dots, d$. Recall that the target density has the form $\pi(x) = \exp(-U(x))/Z$. Let  $(a)_+ = \max(a,0)$ denote the positive part of $a \in \R$. 

We define
\begin{equation} \label{eq:ZZP-rates} \lambda_i(x,v) = (v_i \partial_{x_i} U(x))_+ + \gamma_i(x),\end{equation}
where it should be noted that the \emph{excess switching rate} $\gamma_i(x) \ge 0$ does not depend on $v$.
It is possible to have slightly more general forms of $\gamma_i$, but the current set-up suffices for the purposes of this survey. Often, we simply choose $\gamma_i(x) = \gamma_i \ge 0$ to be a constant, and usually we choose $\gamma_i = 0$.

The jumping mechanism can be interpreted as $d$ Poisson-clocks running synchronously. When the $i$th clock `rings', the $i$th coordinate of $v$ is flipped. This corresponds to the deterministic Markov transition kernel $Q_i$ described by
\[ Q_i(x,v, dv') = \delta_{F_i v}(dv'),\]
where $\delta_w$ represents a Dirac measure centered on $w \in \mathcal V$, and $F_i v$ represents the operation of flipping the $i$th coordinate of $v$:
\[ (F_i v)_j = \begin{cases} v_j, \quad j \ne i, \\
	- v_i, \quad j = i. \end{cases}\]

We can encode the mechanism described above as a single jump rate and transition kernel.
The total jump rate is  then given by $\lambda(x,v) = \sum_{i=1}^d \lambda_i(x,v)$, and the combined jump kernel is
\[ Q(x,v,dv') = \sum_{i=1}^d \frac{\lambda_i(x,v)}{\lambda(x,v)} Q_i(x,v, dv').\]
This fully specifies the Zig-Zag process for a given target density $\pi(x)$.

\subsection{Bouncy Particle Sampler}

The Bouncy Particle Sampler (BPS, \cite{BouchardCoteVollmerDoucet2017}) has a continuous set of velocities $\mathcal V = \R^d$ or, if we wish to work with a compact space of velocities, $\mathcal V = S^{d-1} := \{ v \in \R^d : \|v\| =  1\} $, the unit sphere in $\R^d$.

There are two competing jump mechanisms: a possible ``bounce'', which is informed by the gradient of the target distribution, and a ``refreshment'', which ensures ergodicity by randomly refreshing the velocity at intermittent times.

The bouncing intensity is given by
\[ \lambda_b(x,v) = (\langle v, \nabla U(x) \rangle)_+,\]
and deterministic jump kernel, concentrated at $R(x)v$:
\[ Q_b(x,v, dv') = \delta_{R(x)v} (dv').\]
Here $R(x) : \mathcal V \rightarrow \mathcal V$ denotes the reflection with respect to the contour of $U(x)$. More specifically,
\[ R(x) v = v - 2 \frac{\langle v, \nabla U(x)\rangle}{\|\nabla U(x)\|^2} \nabla U(x).\]
It is easily checked that $\|R(x) v\| = \| v\|$ so that indeed $R(x)$ maps $\mathcal V$ into itself.

The refreshment intensity is given by
\[ \lambda_r (x,v) = \lambda_r(x),\]
that is independent of the velocity $v$ (often chosen to be a positive constant). A refreshment event consists of a random resampling of the velocity, 
\[ Q_r (x,v,dv') = \rho(dv'),\]
where $\rho$ is given by a standard normal distribution if $\mathcal V = \R^d$, and $\rho$ is the uniform distribution on $\mathcal V$  if $\mathcal V = S^{d-1}$.

Again the combination of the different (bounce and refreshment) events can be encoded by a single jump rate
\[ \lambda(x,v) = \lambda_b(x,v) + \lambda_r(x),\]
and jump kernel
\[ Q(x,v, dv') = \frac{\lambda_b(x,v)}{\lambda(x,v)} Q_b(x,v, dv') + \frac{\lambda_r(x)}{\lambda(x,v)} \rho(dv').\] 

\subsection{Stationary distribution}

Having described the construction of the process $(X_t, V_t)_{t \ge 0}$, we now introduce the associated Markov transition kernels,
 \[ P_t ((x,v), A) = \P_{x,v}((X_t, V_t) \in A), \quad A \in \mathcal E,\]
where $\P_{x,v}$ denotes the law of $(X_t, V_t)_{t \ge 0}$ for the given initial value $(x,v)$. Also we write
\[ P_t f(x,v) = \E_{x,v} [f(X_t, V_t)] \quad \text{and} \quad \pi(f) = \int f(x,v) \, \pi (d x, dv).\]
We define the \emph{extended generator} to be the operator $L$ with domain $\mathcal D(L)$ consisting of  functions $f : E \rightarrow \R$ for which 
\[ f(X_t,V_t) - f(x,v) - \int_0^t L f(X_s, V_s)\, d s \]
is a local martingale. 
The domain $\mathcal D(L)$ of $L$ can be characterized explicitly for PDMPs; see \cite[Theorem 26.14]{Davis1993}, and the generator admits the general expression
\begin{equation} \label{eq:generator} L f(x,v) = \langle v, \nabla_x f(x,v) \rangle + \lambda(x,v) (Q f(x,v) - f(x,v)), \quad f \in \mathcal D(L).\end{equation}

 Often we wish to work with a \emph{core}  of the generator, which is a subset of $\mathcal D(L)$ containing `nice' functions, but still sufficiently rich so that properties of the process, such as its stationary distribution, can be determined by considering the generator restricted to $\mathcal C$; see, e.g.,  \cite{durmusPiecewiseDeterministicMarkov2021,Holderrieth2021} for a precise definition.
%
%
% \cite{durmusPiecewiseDeterministicMarkov2021, Holderrieth2021}.
%However with some work a \emph{core} $\mathcal C$ of $L$ can be identified, where $L$ is considered as an unbounded operator on $C_0(E)$, the space of continuous functions on $E$ which vanish at infinity. Such a core 
%
%\todo{the following should be moved}
In the case of the BPS and the ZZP a core can be identified follows.
\begin{proposition} \label{prop:core}
	\begin{itemize}
		\item[(i)] Suppose $U \in C^2(\R^d)$, $\lambda_r > 0$ and $\mathcal V = S^{d-1}$. Then the generator of the BPS admits the core
		\[ \mathcal C = \{ f \in C^1(E) : \text{$f$ has compact support}\}.\]
		\item[(ii)] The generator of the ZZP admits the core
		\[ \mathcal C = \{ f \in C(E) : \text{$f(\cdot, v) \in C^{\infty}(\R^d)$ for all $v \in \mathcal V$ and $f$ has compact support}\}.\]
	\end{itemize}
\end{proposition}
For the proof of parts (i) and (ii) of Proposition~\ref{prop:core} we refer to \cite[Section 9]{durmusPiecewiseDeterministicMarkov2021} and \cite[Section 6.2]{Holderrieth2021}, respectively.

Once the core has been characterized, an integration by parts argument yields the result that BPS and the ZZP admit $\pi$ as  stationary distribution; see \cite{BouchardCoteVollmerDoucet2017, BierkensFearnheadRoberts2016} for a formal derivation.

\section{Ergodicity and exponential ergodicity}
\label{sec:ergodicity}

In \cite{BierkensRobertsZitt2019} it was established that the ZZP is ergodic even when $\gamma_i = 0$ for all $i$. The BPS will in general be reducible when $\lambda_r = 0$ \cite{BouchardCoteVollmerDoucet2017}, with trajectories essentially `circling around the mode', but the BPS will be ergodic when $\lambda_r$ is chosen to be a positive constant.

Under these condition the process $(X_t, V_t)$ is ergodic, and has stationary distribution $\pi(dx,dv)$ with marginal distribution $\pi(dx)$. In particular,  for $\pi$-almost any initial condition,
\[ \mathrm{Law}(X_t,V_t) \rightarrow \pi, \quad t \rightarrow \infty,\]
with convergence in the total variation norm and moreover, the ergodic law of large numbers holds:
\[ \frac 1 T \int_0^T f(X_t, V_t) \rightarrow \int_{\R^d \times \mathcal V} f(x,v)\,  \pi(dx,dv) \quad \text{almost surely as $T \rightarrow \infty$}\]
whenever $f \in L^1(\pi)$ and the left-hand side is well-defined for all $T > 0$.

Under additional assumptions on $U$, exponential ergodicity can be  established for the ZZP \cite{BierkensRobertsZitt2019} and BPS \cite{Deligiannidis2019, DurmusGuillinMonmarche2020}.
%For a given measurable function $V : E \rightarrow [1, \infty)$,  we define a norm $\|\cdot\|_V$ on the space $\mathcal M(E)$ of signed measures on $E$ by
%\[ \|\mu\|_V = \sup_{|f| \le V} |\mu(f)|.\]
%Note that for the total variation norm $[ \|\mu\|_{\mathrm{TV}} = \sup_{|f|\le 1} |\mu(f)|$ 
%we have that  $\|\mu\|_{\mathrm{TV}} \le \|\mu\|_V$ for all $\mu$.
%denotes the \emph{total variation norm}.
We call a continuous time Markov process  with transition kernels $(P_t)_{t \ge 0}$ \emph{$V$-exponentially ergodic}, for a given function $V : E \rightarrow [1,\infty)$, if there are $C > 0$ and $\gamma > 0$ such that for all $f : E \rightarrow \R$ satisfying $|f(z)| \le V(z)$, we have 
\[ | P_t f(z)  - \pi(f)| \le C V(z) \exp(-\gamma t), \quad z \in E. \]

The following result summarizes some findings on exponential ergodicity of BPS and ZZP.
\begin{theorem}[Exponential ergodicity] \label{thm:exponential-ergodicity}
	Suppose $U$ is twice continuously differentiable, $\int_{\R^d} e^{-U(x)}\, d x < \infty$ and $\lim_{\|x\| \rightarrow \infty} U(x) = + \infty$.
\begin{itemize}
	\item[(i)] Suppose $\mathcal V =  S^{d-1}$. Suppose furthermore the following conditions are satisfied:
	\begin{itemize}
		\item[(a)] $\int_{\R^d} \|\nabla U(x)\| e^{-U(x)} \, d x < \infty$;
		\item[(b)] $\int_{\R^d} e^{-U(x)/2} \, d x < \infty$;
		\item[(c)] $\lim_{\|x\| \rightarrow \infty} \|\nabla U(x)\| = \infty$;
		\item[(d)] $\sup_{x \in \R^d} \|\nabla^2 U(x)\| < \infty$.
	\end{itemize}
	then for any choice of constant refreshment rate $\lambda_r > 0$ there exists a $\kappa \in (0,1)$ such that the BPS is $V$-exponentially ergodic for $V = \exp( \kappa U(x))$. 
\item[(ii)] Suppose $\mathcal V= \{-1,+1\}^d$. Suppose furthermore the following conditions are satisfied:
\begin{itemize}
	\item[(a)] $\lim_{\|x\| \rightarrow \infty} \frac{\max(1, \|\nabla^2 U(x)\|)}{\|\nabla U(x)\|} = 0$
	\item[(b)] $\lim_{\|x\| \rightarrow \infty} \frac{\|\nabla U(x)\|}{U(x)} = 0$.
\end{itemize}
Then there exists a function $V \ge 1$ such that the ZZP is $V$-exponentially ergodic.
\end{itemize}
\end{theorem}

\begin{remark}
\begin{itemize}
	\item[(i)] The result for the BPS is from \cite{DurmusGuillinMonmarche2020}; this work also presents (stronger) conditions for which the BPS with $\mathcal V= \R^d$ is exponentially ergodic.
	\item[(ii)] The result for the ZZP is from \cite{BierkensRobertsZitt2019}; the expression for the Lyapunov function is explicit but slightly more complicated than the Lyapunov function for the BPS. Remarkably, the ZZP is (exponentially) ergodic without an additional refreshment.
\end{itemize}
\end{remark}

\section{Central Limit Theorems}
\label{sec:CLT}

Since we have a law of large numbers in some generality, it is natural to ask whether also a Central Limit Theorem holds. The general theory of continuous time Markov processes states that a Functional Central Limit Theorem (FCLT) holds provided we have exponential ergodicity \cite[Theorem 4.3]{GlynnMeyn1996}. This concerns the convergence in distribution of the process defined, for $f : E \rightarrow \R$, and $\overline f(z) := f(z) - \pi(f)$, by 
\[ Y_n(t) = \frac 1 {\sqrt n} \int_0^{nt} \overline f(Z_s)  \, d s, \quad n \in \N, \quad t \ge 0, \]
to a Brownian motion.

\begin{theorem}
Under the assumptions of Theorem~\ref{thm:exponential-ergodicity}, for any function $f : E \rightarrow \R$ such that $|f| \le V$, there is a constant $\sigma_f \ge 0$ such that
\[ Y_n(t) \stackrel{d}{\rightarrow} \sigma_f B,\]
where $B$ is a standard Brownian motion, with convergence in $D[0,1]$, the Skorohod space of cadlag paths.
\end{theorem}

The Functional Central Limit Theorem implies the CLT in the sense that
\[ \frac 1 {\sqrt{n}} \int_0^n \overline f(Z_s)\, d s \stackrel{d}{\rightarrow} N(0,\sigma_f^2).\]
Consequently the CLT provides a quantitative insight in the convergence of ergodic averages through the value of the \emph{asymptotic variance} $\sigma_f^2$. The general theory furthermore states that  $\sigma_f^2$ can be expressed as 
\[ \sigma_f^2 = 2 \int  h(z) \overline f(z) \pi(dz),\]
where $h$ solves the Poisson equation
\[ L h(z) = - \overline f(z), \quad z\in E,\]
with $L$ indicating the generator~\eqref{eq:generator}.

For the one-dimensional Zig-Zag Process, a more explicit expression for $\sigma_f^2$ can be obtained.

\begin{theorem}[{\cite[Theorem 2]{BierkensDuncan2016}}]
For $|f| \le V$, define 
\[ \psi_f(x) := \frac 1 { 2 \pi(x)} \int_x^{\infty} \left( \overline f(\xi,1) + \overline f(\xi,-1) \right) \pi(\xi) \, d \xi. \]
The asymptotic variance associated to $f$ and the ZZP with switching intensity $\lambda$ can be expressed as
\[ \sigma_f^2 = 4 \int_E\lambda (x,v) \psi_f^2(x) d \pi(x,v).\]
\end{theorem}

It can be observed from this expression that increasing $\lambda$ (by increasing the excess switching intensity $\gamma$ in~\eqref{eq:ZZP-rates}) will lead to an increased asymptotic variance $\sigma_f^2$ and thus a slower (asymptotic) convergence of ergodic averages. This observation is further supported by a study of the large deviations from equilibrium for the Zig-Zag Process \cite{BiNySc2021}.

\section{Scaling with dimension}
\label{sec:dimension}

We will now embark on the exploration of various scaling results for BPS and the ZZP in order to obtain a quantitative handle on their sampling properties. The use of scaling limits in the analysis of Markov Chain Monte Carlo sampling algorithms was pioneered in \cite{RobertsGelmanGilks1997}. The goal is to obtain an asymptotic description of the process as some key quantity (such as dimension, or level of anisotropy) approaches infinity. We may have to rescale the time variable in order for a non-trivial limit to exist, and this provides quantitative information on the scaling of the process as a function of the particular quantity of interest.

First we explore the dependence on dimensionality; we refer to \cite{BierkensKamataniRoberts2022} for details.
Consider a Gaussian target distribution in $\R^d$, i.e.
\[ U(x) = \tfrac 1 2 \|x\|^2, \quad x \in \R^d.\] Let $(X^d_t, V^d_t)$ denote the trajectories of the associated PDMPs started in their stationary distribution $\pi$. We consider two derived processes: the \emph{first coordinate} process $C_t^d := (X^{d,1}_t)_{t \ge 0}$ and the (suitably rescaled) \emph{log density} process, $L_t^d := d^{1/2} (d^{-1} \|X^d_t\|^2 - 1)$.
In the following theorem, we understand convergence of the respective one-dimensional processes to mean convergence in distribution under the Skorohod topology.

\begin{theorem}
\begin{itemize}
	\item[(i)] Suppose $\mathcal V = S^{d-1}$. For the BPS we have that, as $d \rightarrow \infty$,
	\begin{itemize}
	\item[(a)] $C_{dt}^d$, i.e., the first coordinate process sped-up by a factor $d$,  converges to the Ornstein-Uhlenbeck process
	\[d X_t = -\lambda_r^{-1} X_t \, d t + \sqrt{2 \lambda_r^{-1}} \, d W_t,\]
	where $(W_t)_{t \ge 0}$ is a standard Brownian motion.
	\item[(b)] $L_{dt}^d$, i.e., the log-density process sped-up by factor $d$, converges to the Ornstein-Uhlenbeck process
	\[ d Y_t = - \frac{\sigma^2(\lambda_r)}{4} Y_t \, d t + \sigma(\lambda_r) \, d W_t,\]
	where $(W_t)_{t \ge 0}$ is a standard Brownian motion, for a known function $\sigma(\lambda_r) > 0$ for $\lambda_r \in (0,\infty)$, satisfying
	\[ \lim_{\lambda_r \downarrow 0} \sigma(\lambda_r) = 0, \quad \lim_{\lambda_r \rightarrow \infty} \sigma(\lambda_r) = 0,\]
	depicted in Figure~\ref{fig:bps-refresh-dependence}.
	\end{itemize}
\begin{figure}[ht!]
	
	\begin{center}	\includegraphics[width = 0.6 \textwidth]{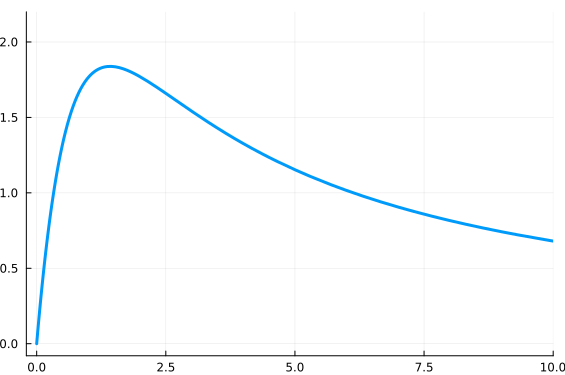}
		\caption{$\sigma^2(\lambda_r)$ as a function of $\lambda_r$. The maximum is attained at $\lambda_r = 1.424$.}
		\label{fig:bps-refresh-dependence}
	\end{center}
\end{figure}
\end{itemize}
\item[(ii)] For the ZZP, with $\mathcal V= \{-1,+1\}^d$, we have that, 
\begin{itemize} \item[(a)] for all values of $d$, the first coordinate process $C_t^d $ is a one-dimensional Zig-Zag process.
\item[(b)] the log-density process $L_t^d$ converges to a non-Markovian stationary Gaussian process with a.s. differentiable paths.
\end{itemize}
\end{theorem}

This theorem yields the following heuristic conclusions. The limiting process of the single coordinate moves at unit speed for both BPS and ZZP. Also the log-density process of the ZZP moves at unit speed in the limit. For the BPS however, the log-density process requires a speed-up by a factor $d$ in order to move at  unit speed (in order of magnitude). The speed $\sigma^2(\lambda_r)$ admits a semi-analytical expression in terms of an  integral, and depends crucially on the value of $\lambda_r$. It can be computed numerically that this speed is optimal for $\lambda_r = 1.424$, and perhaps more importantly, the (scale invariant) ratio of expected number of refreshment jumps to overall jumps is 0.7812. The latter observation can be used more generally as a suitable criterion for tuning of the refreshment rate.

In an analysis as above we should also consider the computational effort required to simulate one continuous time unit of the PDMC sampler. For the ZZP this is $O(d)$ (since every coordinate of the ZZP moves independently at unit speed, and each jump requires changing only one coordinate). For the BPS it turns out there are $O(1)$ velocity jumps, but each jump changes the entire velocity vector, an operation of $O(d)$.

It should be noted that the ZZP complexity per event of $O(1)$ relies heavily on an implementation which utilizes the independence of the ZZP in each direction for this artificial example. An implementation of the ZZP which does not depend on the assumption that every coordinate moves independently, would yield a complexity per event of $O(d)$.

The resulting conclusion for the dimension dependence is as depicted in Table~\ref{tab:dimension-dependence}. For comparison, similar results for classical discrete time MCMC algorithms Random Walk Metropolis (RWM) and Metropolis-Adjusted Langevin Algorithm (MALA) are provided in Table~\ref{tab:dimension-dependence-classical}.

\begin{table}[ht!]
	\small
\begin{tabular}{l|c|c|c|c}
& Time required & Events / unit time & Complexity / event & Complexity \\ 
\hline \hline  
BPS -  First coordinate & $d$ & 1 & $d$ & $d^2$ \\
BPS - Log target & $d$ & 1 & $d$ & $d^2$ \\
\hline
ZZP - First coordinate & 1 & $d$ & $1$ or $d$ & $d$ or $d^2$\\
ZZP - Log target  & 1 & $d$ & $1$ or $d$ & $d$ or $d^2$
\end{tabular}
\caption{Computational complexity to obtain an approximately independent sample for BPS and ZZP as a function of dimension. The final column is obtained by multiplying the other columns. }
\label{tab:dimension-dependence}
\end{table}

\begin{table}[ht!]
	\small
	\begin{tabular}{l|c|c|c}
		& Iterations required & Complexity / sample & Complexity \\ 
		\hline \hline  
		RWM -  First coordinate \cite{RobertsGelmanGilks1997} & $d$ & $d$ & $d^2$ \\
		MALA - First coordinate \cite{RobertsRosenthal1998} & $d^{1/3}$ & $d$ & $d^{4/3}$\\
	\end{tabular}
	\caption{Computational complexity to obtain an approximately independent sample for  classical discrete time algorithms MCMC algorithms RWM and MALA as a function of dimension. The final column is obtained by multiplying the first two columns. }
	\label{tab:dimension-dependence-classical}
	\end{table}
	
A similar dependence of the ZZP on the condition number of a target distribution in the sense of convergence in $L^2(\pi)$ was obtained using the hypocoercivity approach of \cite{Dolbeault2015} in \cite{andrieuHypocoercivityPiecewiseDeterministic2021}.

\subsection{Discussion}
A scaling limit of BPS in a high-dimensional regime was also investigated in \cite{deligiannidisRandomizedHamiltonianMonte2021}. There it is established that, with autoregressive refreshments, the first coordinate process of both position and velocity $(X_t^{d,1},V_t^{d,1})$ of BPS converges weakly to \emph{Randomized Hamiltonian Monte Carlo}.
A notable difference with \cite{BierkensKamataniRoberts2022} is that in \cite{deligiannidisRandomizedHamiltonianMonte2021} the refreshment rate is a factor $O(d^{1/2})$ smaller than the `bounce' rate (i.e., the event rate determined by the potential function).
%For this choice of refreshment rate the theory of \cite{BierkensKamataniRoberts2022} predicts that the log-density process does not converge without additional speeding up.

In \cite{shibaDiffusiveScalingLimits2026} the authors establish similar high-dimensional scaling limits of Forward Event Chain Monte Carlo, a PDMC algorithm similar to BPS but with a different reflection mechanism.

\section{Scaling under anisotropy}
\label{sec:anisotropy}

An alternative setting of interest, studied in \cite{BiKaRo2025}, concerns the case of an anisotropic target distribution where the potential is of the form
\[ U_{\theta,\epsilon}(x) = \tfrac 1 2 x^T  R(-\theta)\Lambda_{\epsilon}^{-1} R(\theta) x, \quad \text{where} \quad R(\theta) = \begin{pmatrix}\cos \theta & - \sin \theta \\ \sin \theta & \cos \theta \end{pmatrix} \quad \text{and} \quad \Lambda_\epsilon =  \begin{pmatrix}1 & 0 \\ 0 & \epsilon \end{pmatrix},\]
i.e., a Gaussian with covariance matrix $\Lambda_\epsilon$ relative to a coordinate system rotated by an angle $\theta$.

We will discuss how BPS and ZZ perform in the asymptotic regime where $\epsilon \downarrow 0$, when the distribution becomes steeply ridged, corresponding to statistical situations of non-identifiability, and, in case $\theta \ne 0$, near-perfect correlation between parameters.

To understand the sampling behaviour, first note that the target distribution is isotropic in a rotated and rescaled coordinate system, described by 
$y^{\epsilon,\theta} = \Lambda_\epsilon^{-1/2} R(\theta)x$. It is easiest to understand this in the case $\theta = 0$, but the other situations are simply rotations of this scenario. We see that the first coordinate of $y$ represents the direction of the target distribution which is of unit magnitude, whereas the other coordinate represents the second direction which becomes infinitesimally concentrated in the limit as $\epsilon \downarrow 0$. Since for BPS and ZZP both coordinates move at unit speed, these small coordinates are oscillating at a high frequency. Consequently this a problem of multiple time scales: a fast time scale for $y_2$  and a slow time scale for $y_1$. It turns out that $y_2$ quickly converges to its stationary distribution.

Write $Y_t^{\epsilon,\theta} :=  \Lambda_\epsilon^{-1/2} R(\theta)X_t$ for the corresponding PDMP (ZZP or BPS) under the target distribution proportional to $\exp(-U_{\theta,\epsilon}(x))$.

In this setting we can phrase the following result.

\begin{theorem} \label{thm:anisotropic-scaling}
	\begin{itemize}
		\item[(i)] Suppose $\mathcal V = \R^d$. For the BPS we have that, as $\epsilon \downarrow 0$, $(Y_t^{1,\epsilon,\theta}, V^1_t)_{t \ge 0}$, converges in distribution to the solution of a certain ODE (with intermittent jumps at refreshment times).
		\item[(ii)] For the ZZP, with $\mathcal V= \{-1,+1\}^d$, we have that, for $\theta \ne k \pi /4$, $k \in \Z$, and as $\epsilon \downarrow 0$, the time-rescaled process
		$(Y_{\epsilon^{-1}t}^{1,\epsilon,\theta})_{t \ge 0}$ converges to the OU-process
		\[ d Y_t = - \frac {\Omega_{\theta}} 2  Y_t \, d t + \sqrt{\Omega_{\theta}} \, d W_t,\]
		where $\Omega_{\theta}$ admits an explicit (but complicated) expression (see Figure~\ref{fig:omega_theta}).
	\end{itemize}
\begin{figure}[ht!]\begin{center}
	\includegraphics[width=0.6 \textwidth]{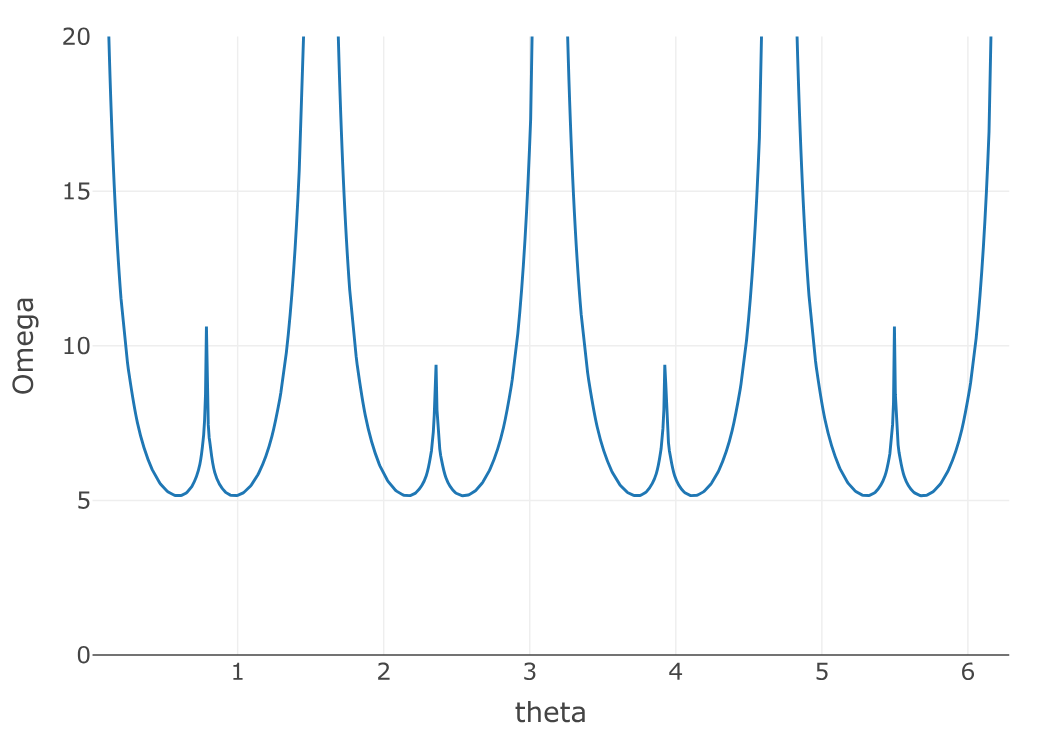}
	\caption{The speed $\Omega_\theta$ of the diffusion scaling limit of the slow coordinate of the Zig-Zag process as a function of the angle $\theta$ of the anisotropic distribution with the $x$-axis.}
	\label{fig:omega_theta}\end{center}
\end{figure}
\end{theorem}

The main conclusions we can draw from Theorem~\ref{thm:anisotropic-scaling} are the following. First, BPS does not require rescaling of time to attain its scaling limit. This is in contrast to the ZZP, which requires speeding-up by a factor $\epsilon^{-1}$. Since the number of switching events per unit time of both processes scales as $\epsilon^{-1}$, it follows that BPS requires a computational effort (interpreted as the number of switching events) of $O(\epsilon^{-1})$ to generate an approximately independent sample, whereas ZZP requires $O(\epsilon^{-2})$ switching events. This result can be compared to an asymptotic complexity of $O(\epsilon^{-2})$ for RWM \cite{beskos2018}.

The result for BPS holds for all $\theta$ since the formulation of BPS is rotation-invariant. For ZZP the speed depends on $\theta$, and for the values $\theta = k \pi/4$, $k \in \Z$, which are not mentioned in the theorem, the convergence is up to an order of magnitude faster as verified by numerical experiment. For even values of $k$ this is easily understood since the two coordinate processes of the ZZP are independent in this case; whereas for odd values of $k$ the correct form of the scaling limit is not yet fully understood.

\section{Scaling in the big data regime}
\label{sec:big-data}

One of the remarkable features of the use of PDMPs is that we can use a stochastic gradient $\widehat{ \nabla U(x)}$ as a replacement for the true gradient $\nabla U(x)$, without losing correctness of the sampling method, in the sense that the PDMP still has $e^{-U(x)}$ as a stationary distribution.

A brief introduction to this idea is as follows. Practical simulation of a PDMP depends upon a Poisson thinning procedure, in which we propose possible switches according to  a tractable upper bound of the true switching rate, and then accept the proposed switch with a certain probability in order to achieve that the switching times follow the desired true rate $\lambda(x,v)$.

It is at the time of this `accept/reject' step that we require evaluation of $\nabla U(x)$, and at this time we can choose to replace it by a stochastic version of the gradient $\widehat {\nabla U(x)}$. Remarkably, if we choose this stochastic gradient in such a way that it is an unbiased estimate for the true gradient, we still have an `effective switching rate' of the correct form~\eqref{eq:ZZP-rates}, with $\gamma_i(x)$ absorbing the randomness in the switching rate induced by the subsampling approach.
For further details of this procedure we refer to  \cite{BierkensFearnheadRoberts2016,fearnhead2018}. In summary, we can use stochastic gradients without affecting the stationary distribution of the PDMC algorithm. However, the trajectories of the algorithm change in way that affects the computational performance, as was investigated in detail in \cite{agrawalLargeSampleScaling2024}; the results of which are briefly summarized below.

\subsection{Stochastic switching rates in the Bayesian big data setting}
For the purpose of this review paper, we consider the Bayesian big data setting in which
\[ U^{(n)}(x) = U_0(x) + \sum_{i=1}^n -\log f(y_i;x),\]
where $(f(\cdot;x))_{x \in \R^d} $ is a parametric family generating data $y_1,\dots, y_n$, and $U_0(x)$ is the contribution from the prior.
In the typical case, we can think of each $U_i(x)$ as representing the contribution of one independent observation to the likelihood (in addition to a contribution from the prior). Write $U_i(x) = U_0(x) - n \log f(y_i;x)$, so that
\[ U^{(n)}(x) = \frac 1 n \sum_{i=1}^n U_i(x).\]

We will consider subsampling schemes that only evaluate a randomly chosen single term in the above expression at each simulation step. Such a subsampling rule thus requires a factor $O(n)$ fewer evaluations of $\nabla U_i$ per simulation step, but we should ask ourselves how the PDMP trajectory is affected.

An obvious candidate for the stochastic gradient is obtained by direct subsampling:
\[ \widehat {\nabla U(x)}_{\mathrm{SS}} = \nabla U_I(x),\] where $I \in \{1, \dots, n\}$ is a uniformly distributed random integer.  

A slightly more complex stochastic gradient is obtained from the subsampling stochastic gradient in addition to a control variate with the goal of variance reduction:
\[ \widehat {\nabla U(x)}_{\mathrm{CV}} = \nabla U_I(x) + \nabla U(\hat x^{(n)}) - \nabla U_I(\hat x^{(n)}),\]
where $\hat x^{(n)}$ is a point `close to the mode' of the posterior, such as the maximum likelihood estimator. The idea of this control variate is that, typically, for large data sets, the posterior will concentrate around $\hat x^{(n)}$ (as a consequence of the Bernstein-von Mises theorem; see \cite{Vaart1998}). Thus $\nabla U_i(x)$ is typically evaluated at $x \approx \hat x^{(n)}$ and we see that for such $x$, the stochastic gradient $\widehat {\nabla U(x)}_{\mathrm{CV}}$ is close to the true gradient by a Taylor expansion:
\begin{align*} \widehat {\nabla U(x)}_{\mathrm{CV}} =\nabla U(x) + [\nabla^2 U_I(\hat x^{(n)}) - \nabla^2 U(\hat x^{(n)})](x- \hat x^{(n)}) + O((x-\hat x^{(n)})^2).\end{align*}
The caveat is that, before sampling, we are required to find a point close to the mode $\hat x^{(n)}$ of the target distribution in order to make the control variates approach work.

For simplicity, we consider the one-dimensional case, so that $x \in \R$ and $v \in \{-1,+1\}$.
Let 
\[\lambda^{(n)}_{\mathrm{can}}(x,v) = \left(v \nabla U^{(n)}(x) \right)_+, \quad \hat \lambda^{(n)}_{\mathrm{SS}}(x,v) = \left(v \widehat{\nabla  U(x)}_{\mathrm{SS}}\right)_+ \quad \text{and} \quad \hat \lambda^{(n)}_{\mathrm{CV}}(x,v)  = \left(v \widehat{ \nabla  U(x)}_{\mathrm{CV}}\right)_+\] denote the switching rates estimators associated with the different choices of stochastic gradients, with $\lambda^{(n)}_{\mathrm{can}}$ denoting the canonical switching intensity. 

\begin{figure}
	\centering
	\begin{subfigure}[t]{\linewidth}
		\centering
		\begin{subfigure}[t]{0.49\linewidth}
			\includegraphics[scale = 0.35]{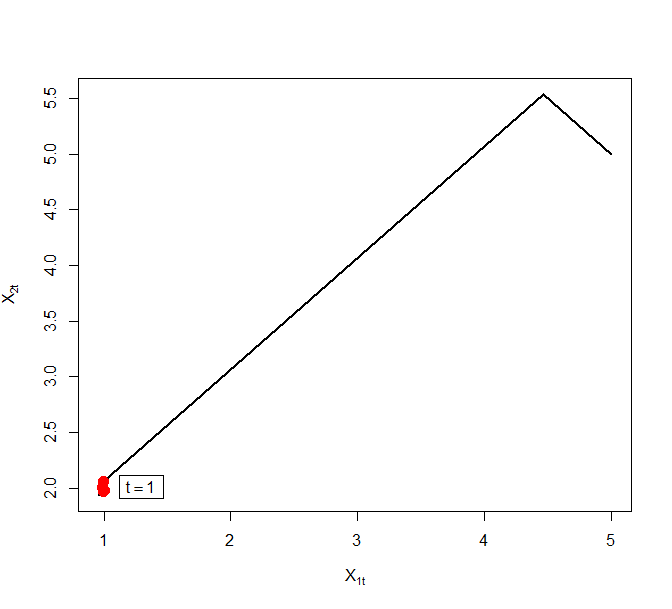}
		\end{subfigure}
		\begin{subfigure}[t]{0.49\linewidth}
			\includegraphics[scale = 0.35]{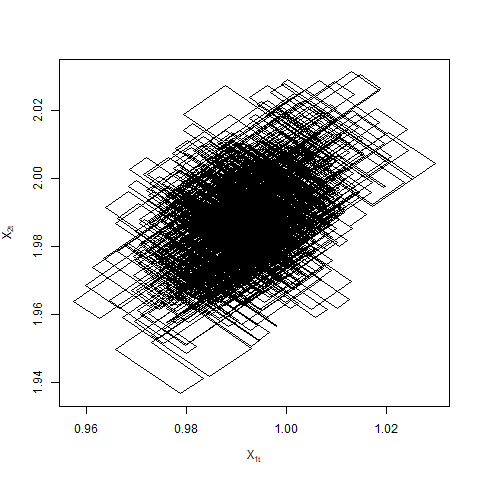}
		\end{subfigure}
		\subcaption{Canonical Zig-Zag (ZZ)}
	\end{subfigure}
	\begin{subfigure}[t]{\linewidth}
		\centering
		\begin{subfigure}[t]{0.49\linewidth}
			\includegraphics[scale = 0.35]{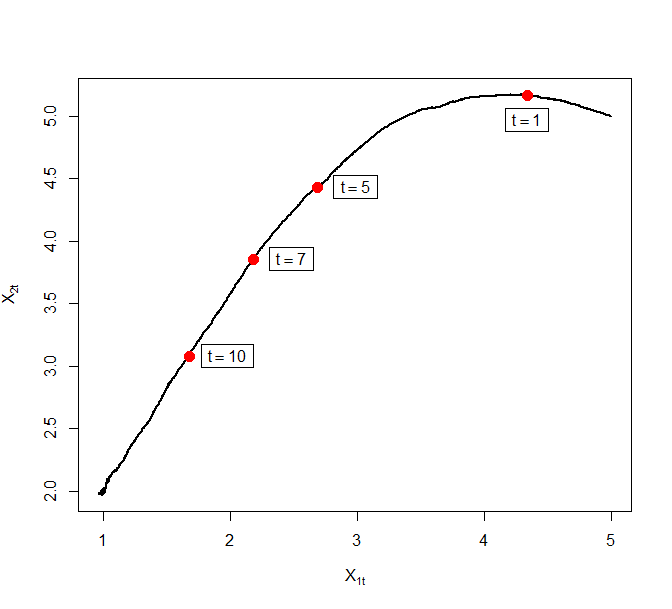}
		\end{subfigure}
		\begin{subfigure}[t]{0.49\linewidth}
			\includegraphics[scale = 0.35]{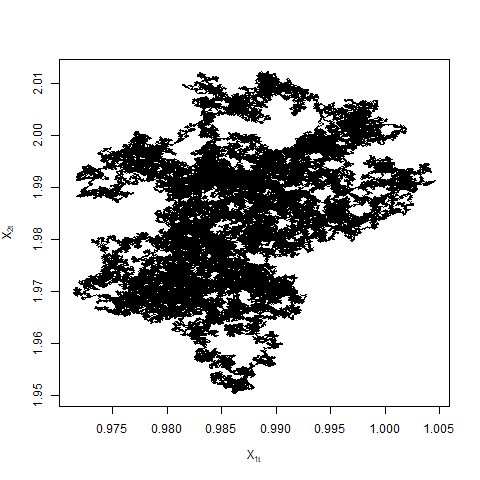}
		\end{subfigure}
		\subcaption{Zig-Zag with vanilla sub-sampling (ZZ-SS)}
	\end{subfigure}
	\begin{subfigure}[t]{\linewidth}
		\centering
		\begin{subfigure}[t]{0.49\linewidth}
			\includegraphics[scale = 0.35]{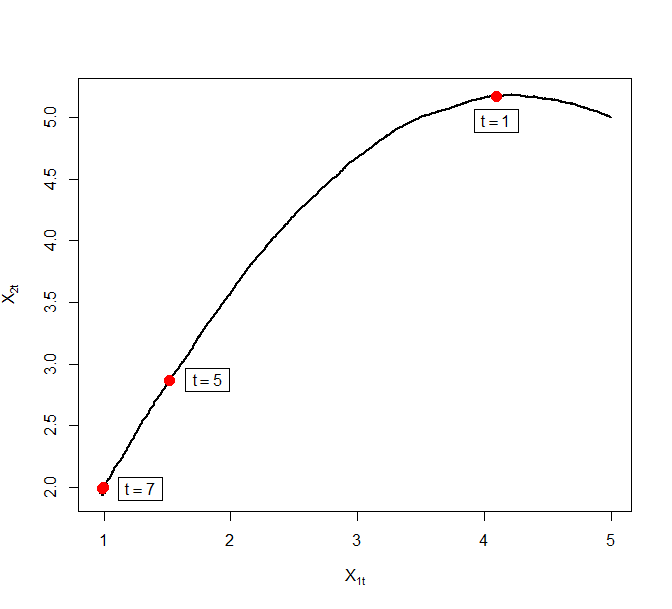}
		\end{subfigure}
		\begin{subfigure}[t]{0.49\linewidth}
			\includegraphics[scale = 0.35]{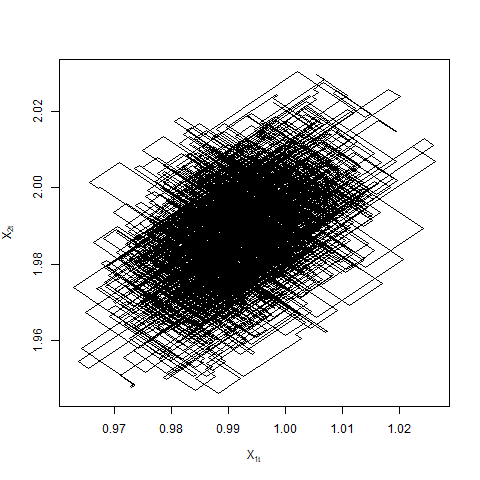}
		\end{subfigure}
		\subcaption{Zig-Zag with control variates (ZZ-CV)}
	\end{subfigure}
	\caption{Trajectories for different versions of the Zig-Zag algorithm targeting a Bayesian posterior for 2-dimensional Bayesian logistic regression with true parameter $(1, 2)$. The left panel displays the trajectories in the transient phase with red points marking the location of the process at different times. The right panel illustrates the process in the stationary phase.}
	\label{fig:subsampling-overview}
	\centering
\end{figure}

We will address two situations of interest: what is the transient behaviour, i.e., when the process is started out of stationarity, and what is the stationary behaviour of the process, as $n \rightarrow \infty$?
Figure~\ref{fig:subsampling-overview} provides a graphical overview of the different subsampling approaches in the transient phase as well as in the stationary phase, which we will now discuss. Mathematical details, including precise rigorous formulations of the mentioned results, can be found in \cite{agrawalLargeSampleScaling2024}.

\subsection{Fluid limits in the transient phase}

Recall that as $n \rightarrow \infty$ the limiting probability distribution $\pi^{(n)}(x) \propto e^{-U^{(n)}(x)}$ is increasingly concentrated around $\hat x^{(n)}$, with a typical distance of $O(n^{-1/2})$, by the Bernstein-von Mises theorem. If we start at a distance $O(1)$ from $\hat x^{(n)}$ we can wonder what the behaviour of the stochastic process will be. 

Let $(X^{(n)}_t, V^{(n)}_t)_{t \ge 0}$ denote the trajectory of the ZZP with one of the switching intensities as described above. In this situation, as $n \rightarrow \infty$ the position process $(X^{(n)}_t)_{t \ge 0}$ converges in distribution to the solution of an ordinary differential equation
\[ x'(t) = b(x(t)), \quad t \ge 0,\]
where the form of the drift $b : \R\rightarrow [-1,1]$ depends on the particular choice of the rate function \cite[Theorem 3.3]{agrawalLargeSampleScaling2024}.

\begin{figure}
	\centering
	\begin{subfigure}{0.325\textwidth}
		\includegraphics[width=\textwidth]{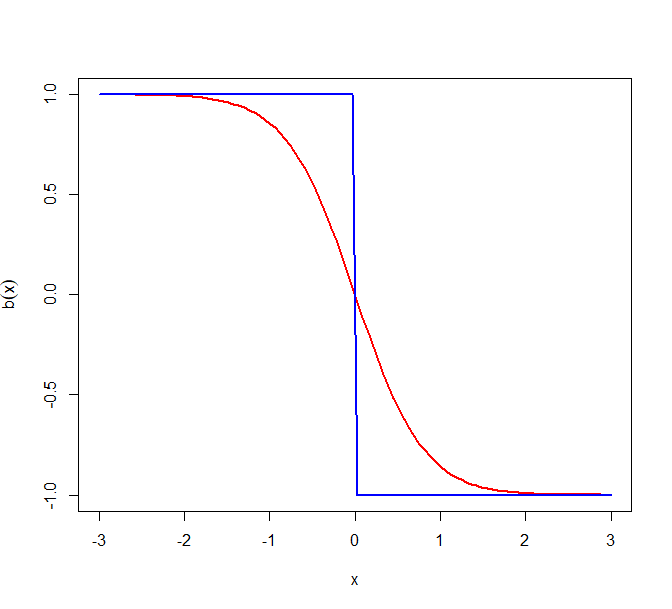}
		\subcaption{Normal}
	\end{subfigure}
	\begin{subfigure}{0.325\textwidth}
		\includegraphics[width=\textwidth]{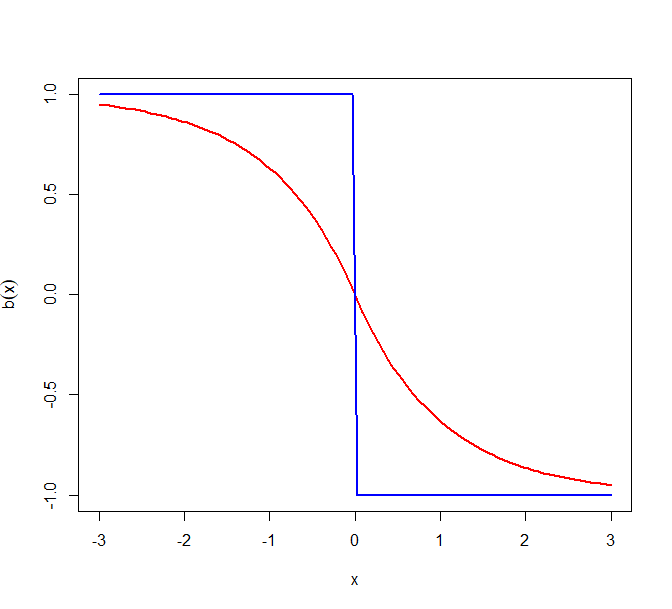}
		\subcaption{Laplace}
	\end{subfigure}
	\begin{subfigure}{0.325\textwidth}
		\includegraphics[width=\textwidth]{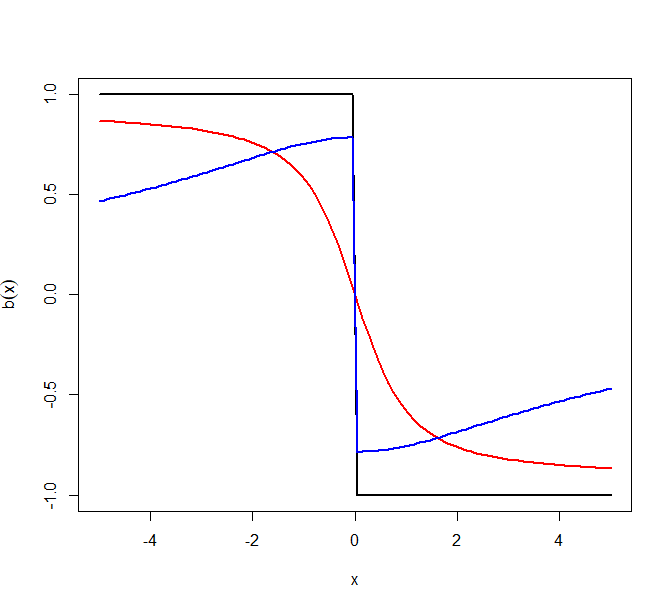}
		\subcaption{Cauchy}
	\end{subfigure}
	\caption{Asymptotic drift $b(x)$ for different models and subsampling schemes in one dimension for three different choices of data generating distribution. In all cases the data is generated from the distribution in the caption with `true' parameter $x = 0$. The black curve denotes the drift corresponding to the canonical rate $\lambda_{\mathrm{can}}$, the red curve corresponds to $\hat \lambda_{\mathrm{SS}}$, and the blue curve corresponds to $\hat \lambda_{\mathrm{CV}}$. Note that in the first two plots, the black and the blue curves overlap.}
	\label{fig:drifts-transient}
\end{figure}

Some examples of the drift functions are provided in Figure~\ref{fig:drifts-transient}. The `ideal' drift function is obtained by using the canonical rate, and corresponds to $b(x) = -\sgn(x)$. This drift gives the fastest drift of $x(t)$ to the mode $x=0$. The control variates approach manages to approximate this canonical drift for the normal distribution and the Laplace distribution. For heavy-tailed data, subsampling provides a better (larger) drift further away from the mode of the target distribution.

\subsection{The stationary phase}

Once we are closer to the mode, say at a distance $O(n^{-1/2})$, the process stops moving as an ODE solution.

As $n \rightarrow \infty$, the process will converge in probability to the constant process at the mode of the distribution (reflecting that the posterior collapses to a Dirac-measure at the mode). However, after rescaling by a factor $n^{1/2}$ we will recover non-trivial behaviour of the limiting processes which we will now describe qualitatively. We refer to \cite{agrawalLargeSampleScaling2024} for details.

In this regime, the canonical Zig-Zag process will exhibit the typical PDMP trajectories. The Zig-Zag process with subsampling (i.e., with stochastic rates $\hat \lambda_{\mathrm{SS}}$) converges to an Ornstein-Uhlenbeck diffusion with speed inversely proportional to the expectation of the absolute value of the stochastic gradient.

The Zig-Zag process with control variates (i.e., with stochastic rates $\hat \lambda_{\mathrm{CV}}$) converges again to a ZZP with a modified rate. In the case where the data arises from an exponential family model, the limiting Zig-Zag process corresponds to the canonical Zig-Zag process.

In other words, ZZP with control variates, generates samples at the same convergence rate as the canonical ZZP, but at a cost reduced by a factor $O(n)$. 

\subsection{Discussion}
The results discussed in this section lead to conclusions as summarized in Table~\ref{tab:summary}; for details see \cite{agrawalLargeSampleScaling2024}.

 \begin{table}
	\caption{Required computational effort for different versions of the Zig-Zag algorithms to obtain an essentially independent sample from stationarity.}
	\label{tab:summary}
	\begin{tabular}{@{}ccccc@{}}
		\hline
		Algorithm & time to an essentially ind. sample & $\sharp$events$/$time & effort/event  & total effort \\
		\hline\\[-2pt]
		ZZ    & $O\left(n^{-1/2}\right)$  & $O\left(n^{1/2}\right)$ & $O(n)$ & $O(n)$ \\
		\\[-2pt]
		ZZ-SS & $O(1)$ & $O(n)$ & $O(1)$ & $O(n)$ \\
		\\[-2pt]
		ZZ-CV & $O\left(n^{-1/2}\right)$  & $O\left(n^{1/2}\right)$ & $O(1)$ & $O(1)$  \\[2pt]
		\hline
	\end{tabular}
\end{table}

\begin{remark}
	To keep the exposition simple, we have not considered stochastic gradients involving batches of a size larger than one. Overall there does not seem to be an advantage to using larger batch sizes: the use of larger batches reduces the variance of the randomized switching rates, but the corresponding gain in efficiency is compensated by the higher computational cost; see \cite{agrawalLargeSampleScaling2024} for details.
\end{remark}

\begin{remark}
	In this section we have discussed in-stationarity and out-of-stationarity (transient) behaviour of the ZZP. A detailed analysis of the transient behaviour of several other PDMC methods (including BPS, the coordinate sampler \cite{WuRobert2018}, and the  Event Chain Monte Carlo sampler \cite{michelForwardEventchainMonte2020a} is provided in the recent work \cite{agrawalTransientRegimePiecewise2025}.
\end{remark}

\subsection*{Acknowledgements}

I am grateful for the collaboration with Sanket Agrawal, Andrew Duncan, Kengo Kamatani and Gareth Roberts in the research projects underlying this survey paper. Much of the challenging rigorous mathematical theory underpinning the scaling results has been established by Sanket Agrawal and Kengo Kamatani. I would also like to thank Sanket Agrawal, Kengo Kamatani and Giorgos Vasdekis for their helpful comments on this survey. The research underlying this survey was for a large part supported by the research programme “Zig-zagging through computational barriers” with project number 016.Vidi.189.043, financed by the Netherlands Organisation for Scientific Research (NWO).

%\bibliographystyle{plain}
%\bibliography{/home/joris/Zotero/library.bib}

\end{document}